\documentclass[aps,amjphys,twocolumn,floatfix,nofootinbib]{revtex4-2}

\usepackage{amsmath, amssymb, amsfonts}
\usepackage{graphicx}
\usepackage[utf8]{inputenc}
\usepackage{hyperref}

\usepackage{pgfplots}
\pgfplotsset{compat=1.18}
\usepackage{tikz}
\usetikzlibrary{shapes.geometric, arrows.meta, positioning, calc}

\newcommand{\dr}{\dot{r}}

\begin{document}

\title{Why does walking to the center of a merry-go-round feel so hard?\\
Coriolis stabilization and the metabolic cost of staying on track}

\author{Mario J. Pinheiro}
\email{mpinheiro@tecnico.ulisboa.pt}
\affiliation{Department of Physics, Instituto Superior T\'ecnico,
Universidade de Lisboa, 1049-001 Lisboa, Portugal}

\date{\today}

\begin{abstract}
We revisit Feynman's classic carousel problem (\textit{Feynman Lectures},
Vol.~1, Sec.~19.4), in which a student walks radially inward on a
platform driven at constant angular velocity $\omega_0$ by an external
motor. Working consistently in the frame co-rotating with the platform,
we show that the student's kinetic energy in that frame is exactly
\emph{constant}: the mechanical work done by the real radial friction
force is exactly cancelled by the work done against the centrifugal
inertial force. The Coriolis inertial force, and the equal-and-opposite
tangential friction force required to cancel it and keep the student on
a straight radial path, do \emph{zero} mechanical work, because both
are perpendicular to the radial velocity. Yet generating that tangential
force costs metabolic energy --- precisely the effect Feynman flagged
with ``one has to lean over and push sidewise.'' The mechanical model
adopted here (a point mass with kinematically imposed trajectory) is
deliberately silent about internal physiology; the metabolic cost
discussion of Sec.~\ref{sec:cost} is a separate phenomenological layer,
explicitly flagged as such, not derived from the mechanics. We give an
order-of-magnitude metabolic estimate, an entropy-production argument
connecting the exercise to the second law, a feedback (PD-controller)
reformulation of the same physics, and a playground experiment students
can run with a phone and a heart-rate strap. Throughout, we are explicit
about reference frames and about the limits of each model.
\end{abstract}

\maketitle

\section{Introduction, and a necessary word about frames}
\label{sec:intro}

Feynman's \textit{Lectures}, Vol.~1, Sec.~19.4, poses a carousel driven
at constant angular velocity $\omega_0$ by a motor, with a person walking
radially on it. Feynman's own remark is the seed of this paper:
``To walk along the radius in a carousel, one has to lean over and push
sidewise. Try it sometime'' \cite{Feynman}. That sideways push is the
Coriolis effect, and it is the subject of this note.

Before any equation is written, three frames must be distinguished,
because the physics looks different --- and some quantities are simply
\emph{not defined} --- in each:

\begin{description}
\item[Lab frame.] An inertial frame in which the carousel's center is
fixed and the platform rotates at $\omega_0$ (held constant by the
motor; this is an external torque doing work on the system, so the
platform--student system is \emph{not} isolated and the student's own
angular momentum is \emph{not} conserved).

\item[Rotating frame.] The non-inertial frame co-rotating with the
platform at $\omega_0$. This is the frame in which centrifugal and
Coriolis inertial forces appear.\footnote{We use the term \emph{inertial force} rather than
\emph{fictitious force} or \emph{pseudoforce}, following a suggestion
by K.~T.~McDonald \cite{McDonaldComment}: the word ``fictitious''
misleadingly implies that only inertial frames are physically valid,
whereas these forces are entirely real to an observer in the rotating
frame. The adjective refers to the frame, not to the physics.} It is the frame in
which ``the idealized walker moves a straight line toward the center'' is the
natural description of the trajectory, and where we do all the mechanics.

\item[Student's rest frame.] The frame in which the student's center of
mass is instantaneously at rest. In this frame $v=0$ for the center of
mass, no force does any work on it, and internal limb motion relative to
the center of mass is the entire story --- a separate problem we do not
attempt here.
\end{description}

All mechanics in this paper is done in the \textbf{rotating frame},
stated explicitly at every step.

\subsection*{Learning objectives}
After working through this material a student should be able to: (i)
state, for a driven carousel, what is and is not conserved in the lab
frame versus the rotating frame, and why; (ii) compute the student's
kinetic energy in the rotating frame and show it is constant for
constant radial speed; (iii) identify which forces do and do not do
work, and derive the Coriolis force explicitly; (iv) build and
critically evaluate a phenomenological metabolic cost model, including
its domain of validity; (v) connect the mechanical exercise to entropy
production and to a feedback-control description.

\section{The mechanical model}
\label{sec:model}

We model the idealized walker as a \emph{point mass} $m$ whose centre of mass
moves radially inward at constant speed $|\dr|=v_0$, with the trajectory
maintained by the forces of contact (friction) between the idealized walker and
the platform. We do not model the internal mechanism --- limbs, muscles,
gait cycle --- that produces this motion. This is a kinematic
imposition, not a dynamical derivation.

This choice is deliberate and has a well-known implication for the work
budget: for any locomotion in which the foot (or contact point) is
instantaneously at rest relative to the supporting surface ---
in models of bipedal walking where the foot is instantaneously at rest
relative to the ground --- including rolling-without-slipping idealizations ---
the friction force of the surface on the moving body does \emph{zero
instantaneous work} \cite{McDonaldComment}. In such models the energy
that drives the motion comes from internal sources. Our point-mass model
does not resolve this internal mechanism; we simply note the consistency. The energy that drives the
motion comes from internal sources (ATP hydrolysis, stored elastic
energy in tendons), not from the work of friction. Minimal models of
bipedal walking that make this explicit exist in the literature
\cite{Katwal2024}; the augmented inverted-pendulum model of
Katwal et al., for example, shows that centre-of-mass kinematics
alone can yield realistic estimates of the mechanical cost of
transport per step, precisely because the constraint forces that
drive the motion are determined by inverse dynamics rather than
assumed. We do not pursue such models here. Our point-mass
model is therefore consistent with this constraint: we impose the
kinematics and ask what forces the rotating environment requires,
deferring all questions about internal energy to
Sec.~\ref{sec:cost}.

We work in plane polar coordinates $(r,\theta)$ in the rotating frame,
defined so that the platform sits at constant $\theta$. The carousel is
driven at $\omega_0=\text{const}$, so the frame's angular velocity
$\boldsymbol{\Omega}=\omega_0\hat{z}$ is constant ($\dot{\boldsymbol{\Omega}}=\mathbf{0}$)
and its origin coincides with the lab-frame origin (no translational
acceleration). Under these conditions the general noninertial equation
of motion \cite{LandauLifshitz} reduces to
\begin{equation}
m\,\mathbf{a}_{\mathrm{rot}}
   = \mathbf{F}_{\mathrm{friction}}
   - m\,\boldsymbol{\Omega}\times(\boldsymbol{\Omega}\times\mathbf{r})
   - 2m\,\boldsymbol{\Omega}\times\mathbf{v}_{\mathrm{rot}},
\label{eq:eom_rotating}
\end{equation}
where $\mathbf{F}_{\mathrm{friction}}$ is the contact force of the
platform on the idealized walker (static friction, both radial and
tangential components), the second term is the centrifugal inertial
force, and the third is the Coriolis inertial force. Here
$\mathbf{a}_{\mathrm{rot}}$ is the acceleration of the \emph{centre
of mass} of the idealized body, not of the limbs of an actual walking
person. The terms associated with a translating or angularly
accelerating frame vanish identically here.

The trajectory of the idealized walker in the rotating frame is purely radial:
$\theta=\theta_0=\text{const}$, $r(t)=r_0-v_0 t$. With
$\dot\theta_{\mathrm{rot}}=0$ and $\ddot\theta_{\mathrm{rot}}=0$,
\begin{equation}
a_r = \ddot r = 0
\quad (\text{constant }v_0),
\qquad a_\theta = 0.
\label{eq:polar_accel}
\end{equation}
Both components of the acceleration vanish, and both must be enforced
by the friction force against the inertial forces that would otherwise
deflect the student.

\section{Kinetic energy in the rotating frame: a constant}
\label{sec:KE}

In the rotating frame the velocity of the idealized walker's centre of mass is purely radial:
$\mathbf{v}_{\mathrm{rot}} = \dr\,\hat{r}$. The kinetic energy is
\begin{equation}
K_{\mathrm{rot}} = \tfrac{1}{2}m\dr^{2} = \tfrac{1}{2}mv_0^{2}
   = \text{const}.
\label{eq:Krot}
\end{equation}
This is constant because $v_0$ is constant by assumption. It is
\emph{not} the result of some special cancellation; it is a direct
consequence of the imposed kinematics. The student's kinetic energy
in the rotating frame neither rises nor falls as the center is
approached.

We note, for completeness, that in a more realistic model of walking
--- one that includes limb motion relative to the centre of mass ---
the kinetic energy in the rotating frame would be time-varying, since
the limbs oscillate even while the centre of mass moves at constant
mean speed. Our point-mass model is silent about this, and
Eq.~(\ref{eq:Krot}) should be read as a statement about the
centre-of-mass kinetic energy only.

\section{Which forces do work}
\label{sec:work}

Equation~(\ref{eq:Krot}) being constant means the net power on the
student is zero. We now identify the individual contributions, as the
correct accounting requires checking every term in
Eq.~(\ref{eq:eom_rotating}) \cite{McDonaldComment}.

\subsection{Centrifugal inertial force}

The centrifugal inertial force is radial, outward:
$\mathbf{F}_{\mathrm{cf}} = m\omega_0^2 r\,\hat{r}$. Since the
student's velocity is radial, the power is
\begin{equation}
P_{\mathrm{cf}} = m\omega_0^2 r\,\dr < 0
\qquad (\dr < 0).
\label{eq:Pcf}
\end{equation}
The centrifugal force does negative work as the idealized walker moves inward.

\subsection{Radial friction force}

From the radial component of Eq.~(\ref{eq:eom_rotating}) with $a_r=0$:
\begin{equation}
F_{r,\mathrm{friction}} = m\omega_0^2 r \quad (\text{inward}).
\label{eq:Fr}
\end{equation}
The platform must push the student inward to counteract the outward
centrifugal force. The power of this friction force is
\begin{equation}
P_{r,\mathrm{friction}} = F_{r,\mathrm{friction}}\,\dr
   = -m\omega_0^2 r\,\dr > 0
\qquad (\dr < 0),
\label{eq:Preal}
\end{equation}
exactly cancelling Eq.~(\ref{eq:Pcf}). This confirms
$dK_{\mathrm{rot}}/dt = 0$.

\subsection{Coriolis inertial force}

With $\boldsymbol{\Omega}=\omega_0\hat{z}$ and
$\mathbf{v}_{\mathrm{rot}}=\dr\,\hat{r}$:
\begin{equation}
\mathbf{F}_{\mathrm{Cor}}
   = -2m\boldsymbol{\Omega}\times\mathbf{v}_{\mathrm{rot}}
   = -2m\omega_0\dr\,\hat\theta,
\qquad
|\mathbf{F}_{\mathrm{Cor}}| = 2m\omega_0 v_0.
\label{eq:Fcor}
\end{equation}
This force is purely tangential. Since the student's velocity is purely
radial, $\mathbf{F}_{\mathrm{Cor}}\cdot\mathbf{v}_{\mathrm{rot}}=0$:
\emph{the Coriolis inertial force does zero work}.

\subsection{Tangential friction force}

From the tangential component of Eq.~(\ref{eq:eom_rotating}) with
$a_\theta=0$:
\begin{equation}
F_{\theta,\mathrm{friction}} = 2m\omega_0 v_0
\quad (\text{tangential, opposing Coriolis}).
\label{eq:Ftheta}
\end{equation}
This is Feynman's ``push sidewise'': the platform exerts a tangential
friction force on the student to prevent lateral drift. Since the
contact point is instantaneously at rest in the rotating frame (the
student does not slide on the platform), this tangential friction force
also does \emph{zero work} in the rotating frame.

\subsection{Summary}

The power balance in the rotating frame is:
\begin{equation}
\underbrace{P_{\mathrm{cf}}}_{<\,0}
+\underbrace{P_{r,\mathrm{friction}}}_{>\,0}
+\underbrace{P_{\mathrm{Cor}}}_{=\,0}
+\underbrace{P_{\theta,\mathrm{friction}}}_{=\,0}
= \frac{dK_{\mathrm{rot}}}{dt} = 0.
\label{eq:powerbal}
\end{equation}
The conceptual core: the tangential friction force
[Eq.~(\ref{eq:Ftheta})] does zero mechanical work on the student's
centre of mass, yet generating it requires metabolic energy. This is
the precise mechanical statement of what Feynman described qualitatively.

\section{The lab-frame energy balance}
\label{sec:labframe}

In the lab frame the student's velocity has radial component $\dr$ and
tangential component $r\omega_0$, giving
\begin{equation}
K_{\mathrm{lab}} = \tfrac{1}{2}m\!\left(\dr^{2}+r^{2}\omega_0^{2}\right),
\label{eq:Klab}
\end{equation}
with
\begin{equation}
\frac{dK_{\mathrm{lab}}}{dt} = -m\omega_0^{2}v_0\,r(t) < 0.
\label{eq:dKlab}
\end{equation}
The student's lab-frame kinetic energy \emph{decreases} as the center
is approached, because the tangential speed $r\omega_0$ falls faster
than any radial contribution grows.

The lab-frame energy budget closes through the motor. In the lab frame,
the contact point between student and platform moves tangentially at
$r\omega_0$, so the tangential friction forces (student on platform, and
platform on student) are an action--reaction pair both acting at a point
with nonzero tangential velocity. However, being equal and opposite,
their total work on the system is zero. The platform's kinetic energy
is constant ($\omega_0$ fixed, $I$ fixed). Applying the work--energy
theorem to the entire system:
\begin{equation}
W_{\mathrm{motor}} = \Delta K_{\mathrm{lab}} =
   -\tfrac{1}{2}m\omega_0^{2}(r_0^{2}-r_f^{2}) < 0.
\label{eq:Wmotor}
\end{equation}
The motor does \emph{negative} work: as the idealized walker moves inward and
the system's effective moment of inertia decreases, the motor must
\emph{absorb} rotational energy at fixed $\omega_0$, acting as a
brake. This result follows without approximation and closes the
lab-frame energy budget completely.

\section{Metabolic cost of Coriolis stabilization}
\label{sec:cost}

\subsection{The gap between mechanics and physiology}
\label{sec:gap}

The sections that follow step deliberately outside the mechanical model
of Secs.~\ref{sec:model}--\ref{sec:labframe}. Their purpose is to
motivate physiological questions and point toward the kind of model
that would be needed to answer them, not to derive metabolic predictions
from the point-mass mechanics above. Quantifying the metabolic cost
rigorously would require a model of walking that includes legs, feet,
and a gait cycle \cite{Katwal2024}; we do not attempt that here.

The mechanical model of Secs.~\ref{sec:model}--\ref{sec:labframe} is
deliberately silent about internal physiology. It tells us that a
tangential friction force of magnitude $F_C = 2m\omega_0 v_0$ must act
on the student at all times, that this force does zero mechanical work
on the centre of mass, and that the energy to sustain the motion comes
from internal sources. It does not tell us how much internal energy is
required to generate $F_C$.

To address that question we must step outside the mechanical model and
introduce a phenomenological hypothesis. The literature distinguishes
at least four types of metabolic expenditure that a single-coordinate
model cannot separate: (i)~\emph{mechanical work on the centre of
mass}; (ii)~\emph{internal muscular work} associated with limb motion
relative to the centre of mass; (iii)~\emph{isometric cost}, the ATP
consumed by cross-bridge cycling to hold tension without macroscopic
displacement --- as when holding a heavy book at arm's length,
stationary: mechanically $W=0$, yet oxygen uptake rises measurably
\cite{Alexander}; and (iv)~\emph{active stabilization cost}, the
metabolic expenditure of generating a force that does zero work on the
centre of mass but is required continuously to maintain a desired
trajectory. Our case is of type~(iv), and possibly also type~(ii) and
(iii). We do not attempt to decompose these contributions further, as
doing so would require a detailed model of gait along the lines of \cite{Katwal2024}
beyond the scope of this note.

\subsection{A family of cost models}
\label{sec:family}

We adopt a family of phenomenological cost laws relating metabolic power
to the magnitude of the force the student must generate:
\begin{equation}
P_{\mathrm{met}} = \alpha_n\,F_C^{\,n}.
\label{eq:family}
\end{equation}
This is a hypothesis about the biology, not a consequence of the
mechanics. Three values of $n$ are useful for discussion: linear
($n=1$, supported by some lumped muscle-cost models \cite{Nelson});
quadratic ($n=2$, the canonical linear-quadratic optimal-control choice
\cite{Astrom}, formally identical to Joule heating, and motivated for
stochastic stabilization tasks by Kuo \cite{Kuo}); and higher order
($n>2$, relevant near the force limit of the muscle \cite{Nelson}).
The number $P_{\mathrm{met}}$ should be read as an order-of-magnitude
anchor, not a biomechanical prediction: $\alpha_n$ is calibrated from
analogous but not identical tasks, as described below.

\subsection{Scaling with radius}
\label{sec:scaling}

Since $F_C = 2m\omega_0 v_0$ is independent of $r$ in the driven
scenario, substituting into Eq.~(\ref{eq:family}) gives
\begin{equation}
P_{\mathrm{met}} = \alpha_n\left(2m\omega_0 v_0\right)^{n}
   = \text{const}.
\label{eq:Pscaling}
\end{equation}
The metabolic cost is \emph{flat} with $r$ for constant $\omega_0$ and
$v_0$ --- a real, sustained cost, but not a diverging one. Whether
effort escalates near the axis depends on scenario: in the isolated
coasting-platform case (Appendix~\ref{app:isolated}), $\omega(r)$ rises
and $F_C$ grows. In the driven case, any escalation comes not from a
growing force but from a shrinking control margin, as discussed in
Sec.~\ref{sec:control}.

\subsection{Order-of-magnitude estimate}
\label{sec:numerics}

Take $m=60$~kg, $v_0=0.5$~m/s, $\omega_0=1$~rad/s. Then
$F_C = 60$~N. For the quadratic model, anchoring to a nominal
$P_{\mathrm{met}}=200$~W in line with stabilization costs reported
for comparable postural tasks \cite{Hogan,Kuo} --- a borrowed, not
task-specific, calibration --- gives
$\alpha_2 = 200/3600 \approx 0.056$~W/N$^2$.
Because $F_C$ is constant, $P_{\mathrm{met}}\approx200$~W throughout
the trip: substantial and physiologically plausible, but flat.

\section{A thermodynamic reading}
\label{sec:thermo}

Skeletal muscle converts chemical free energy to mechanical output with
efficiency $\eta\sim0.25$ \cite{Alexander}. Since the tangential
friction force does zero mechanical work, essentially all of
$P_{\mathrm{met}}$ associated with generating $F_C$ is dissipated as
heat, giving entropy-production rate
\begin{equation}
\dot{S}_{\mathrm{gen}} \approx \frac{P_{\mathrm{met}}}{T}
\approx 0.6~\mathrm{W/K}
\label{eq:entropy}
\end{equation}
at $T\approx310$~K and $P_{\mathrm{met}}\approx200$~W. A passive bead
on a frictionless radial wire would require no tangential force and
generate no entropy (Problem~P3). This contrast captures the essential
thermodynamic difference between passive mechanical stability and active
biological stabilization: the former is free; the latter is
necessarily dissipative.

\section{Feedback control: the same physics, reframed}
\label{sec:control}

The discussion above treated the student as a black box that generates
$F_C$ perfectly. A complementary picture is that the student is a
\emph{controller}: she senses lateral error and commands corrective
force. In the rotating frame the tangential equation of motion,
allowing for imperfect cancellation via a control force $u_\theta(t)$,
is
\begin{equation}
m(r\ddot\theta + 2\dr\dot\theta) = u_\theta(t) + F_{\mathrm{Cor},\theta},
\end{equation}
where $F_{\mathrm{Cor},\theta} = -2m\omega_0\dr$ is the Coriolis
disturbance. A PD controller on the angular tracking error
$e=\theta-\theta_0$,
\begin{equation}
u_\theta = -k_p e - k_d\dot{e},
\label{eq:PD}
\end{equation}
gives closed-loop dynamics
\begin{equation}
mr\ddot{e} + k_d\dot{e} + k_p e = -2m\omega_0\dr,
\label{eq:closed_loop}
\end{equation}
a driven damped oscillator with natural frequency
$\omega_n=\sqrt{k_p/(mr)}$ and damping ratio
$\zeta=k_d/(2\sqrt{mrk_p})$. At fixed gains, $\omega_n$ rises as $r$
falls: the loop becomes harder to stabilize near the center even
though the disturbance amplitude $|F_{\mathrm{Cor}}|=2m\omega_0 v_0$
is constant. This rising natural frequency --- and the tendency toward
underdamping at fixed gains --- is a genuinely testable prediction: the
frequency of corrective lateral sway should grow as $r^{-1/2}$ as the
axis is approached.

\section{A back-of-the-envelope experiment}
\label{sec:experiment}

A playground roundabout, a smartphone (gyroscope and accelerometer),
and a heart-rate strap suffice for a qualitative test. Two predictions
are accessible without laboratory instrumentation: (1) in the driven
scenario, metabolic cost (heart-rate proxy) should be roughly
\emph{flat} with $r$ at constant $\omega_0$ and $v_0$ --- a
falsifiable claim that distinguishes the driven from the coasting
scenario; (2) the dominant frequency of corrective lateral sway should
rise as $r$ decreases, per Sec.~\ref{sec:control}. Neither measurement
is publication-grade, and that is the point: students confront real,
noisy data and decide what the model robustly predicts versus what it
does not. Safety notes --- spotter, low rotation rate, helmet --- belong
in any handout.

\section{Classroom use}
\label{sec:classroom}

We suggest the following sequence: (1) the standard Feynman problem,
establishing $\omega_0=\text{const}$ and the role of the motor; (2) the
mechanical model of Sec.~\ref{sec:model} and the explicit identification
of $\mathbf{F}_{\mathrm{friction}}$; (3) the force-by-force work
analysis of Sec.~\ref{sec:work}, which is the most error-prone step;
(4) the lab-frame cross-check of Sec.~\ref{sec:labframe}; (5) the
explicit separation between mechanical model and phenomenological
metabolic layer (Sec.~\ref{sec:cost}); (6) the entropy argument; and,
time permitting, (7) the control reformulation and experiment.

The recurring meta-lesson is threefold: \emph{(a)} getting the
reference frame right is not a formality --- it determines the entire
mechanical conclusion; \emph{(b)} a force can do zero mechanical work
and still cost metabolic energy; \emph{(c)} the boundary between a
mechanical model and a biological one must be stated explicitly, not
assumed.

\section{Conclusion}
\label{sec:conclusion}

Working in the rotating frame of a motor-driven carousel, we find that
the idealized walker's kinetic energy is exactly constant, that the
centrifugal and radial friction forces trade equal and opposite work
leaving $K$ unchanged, and that the Coriolis inertial force and its
tangential friction counterpart do zero mechanical work on the centre
of mass --- suggesting, in a real biological walker, a possible
metabolic cost associated with active stabilization, which is the
qualitative content of Feynman's remark about leaning and pushing
sidewise. Quantifying that cost rigorously would require a model of
walking that includes legs, feet, and a gait cycle \cite{Katwal2024};
the present paper does not attempt this. The mechanical model is
deliberately minimal --- a point mass with imposed kinematics --- and
the phenomenological sections that follow are a hypothesis about
biology, not a derivation from mechanics. The lab-frame energy budget closes cleanly: the motor
absorbs energy rather than supplying it, acting as a brake as the
system's moment of inertia decreases. We hope the paper is useful both
for the physics it contains and as an example of how to state the limits
of a model as clearly as its predictions.

\begin{acknowledgments}
The author is deeply grateful to Prof.\ Kirk T.\ McDonald (Princeton
University) for two rounds of detailed critique that corrected
frame-consistency errors, identified the role of friction as the real
force in the model, clarified the work done by each force individually,
and pointed to the literature on minimal walking models. The paper is
substantially better for his engagement. Any remaining errors are the
author's own.
\end{acknowledgments}

\appendix

\section{The isolated, coasting-platform scenario}
\label{app:isolated}

For completeness we treat the case in which the total angular momentum
of platform plus student is conserved (no motor; low-friction
bearings):
\begin{equation}
L_{\mathrm{total}} = (I+mr^2)\omega = I\omega_0,
\label{eq:Ltot}
\end{equation}
giving
\begin{equation}
\omega(r) = \frac{I\omega_0}{I+mr^2}.
\label{eq:omega_of_r}
\end{equation}
Here $\omega$ rises as $r$ falls, and the Coriolis force
$F_C(r) = 2mv_0\omega(r)$ increases smoothly but saturates at
$2mv_0\omega_0$ as $r\to0$ --- bounded, not divergent. The metabolic
cost under any cost model therefore also saturates. This case is
arguably the more common playground scenario and gives a rising, bounded
metabolic demand, in contrast to the flat cost of the driven scenario.

\section{Problem set}
\label{app:problems}

\noindent\textbf{P1.} For the motor-driven carousel, show explicitly
that the student's own angular momentum $mr^2\omega_0$ is not conserved
as $r$ decreases, and explain physically (in terms of the external
torque of the motor) why not.

\smallskip\noindent\textbf{P2.} Reproduce the power balance of
Eq.~(\ref{eq:powerbal}): compute $P_{\mathrm{cf}}$,
$P_{r,\mathrm{friction}}$, $P_{\mathrm{Cor}}$, and
$P_{\theta,\mathrm{friction}}$ and verify they sum to zero.

\smallskip\noindent\textbf{P3.} A frictionless bead on a radial wire
rotating with the platform replaces the student. Which forces in
Eq.~(\ref{eq:eom_rotating}) are supplied by the wire? Is any metabolic
energy required? Explain why $\dot{S}_{\mathrm{gen}}=0$ for the bead
but not for the student.

\smallskip\noindent\textbf{P4.} Derive Eq.~(\ref{eq:omega_of_r}) and
verify that $F_C(r)$ saturates rather than diverges as $r\to0$.

\smallskip\noindent\textbf{P5.} Using $\eta\approx0.25$ and $T=310$~K,
estimate the total entropy generated as the student walks from
$r_0=2$~m to $r_f=0.5$~m under the driven scenario, and compare it to
$\Delta K_{\mathrm{lab}}$ from Eq.~(\ref{eq:Wmotor}).

\smallskip\noindent\textbf{P6 (open-ended).} Design the playground
experiment of Sec.~\ref{sec:experiment} to distinguish the driven from
the coasting scenario using heart-rate data alone. What is the key
observable difference?

\end{document}